\documentclass[%
 aip,
 amsmath,amssymb,
 reprint,%
]{revtex4-1}

\usepackage{graphicx}
\usepackage{dcolumn}
\usepackage{bm}

\usepackage[utf8]{inputenc}
\usepackage[T1]{fontenc}
\usepackage{mathptmx}
\usepackage{tikz}
\usepackage{etoolbox}

\newcommand{\Ha}{\mathcal{H}}

\newcommand{\half}{\frac{1}{2}}

\makeatletter
\def\@email#1#2{%
 \endgroup
 \patchcmd{\titleblock@produce}
  {\frontmatter@RRAPformat}
  {\frontmatter@RRAPformat{\produce@RRAP{*#1\href{mailto:#2}{#2}}}\frontmatter@RRAPformat}
  {}{}
}%
\makeatother
\begin{document}

\preprint{AIP/123-QED}

\title[Phonon Amplification via Magnetoelastic Klein Scattering]{Phonon Amplification via Magnetoelastic Klein Scattering}
\author{A.L. Bassant}
 \email{a.l.bassant@uu.nl.}
 \author{M.F. van Willigen}
\affiliation{Institute for Theoretical Physics, Utrecht University, Princetonplein 5, 3584CC Utrecht, The Netherlands}
\author{R.A. Duine}
\affiliation{Institute for Theoretical Physics, Utrecht University, Princetonplein 5, 3584CC Utrecht, The Netherlands}
\affiliation{Department of Applied Physics, Eindhoven University of Technology, P.O. Box 513, 5600 MB Eindhoven, The Netherlands}

\date{\today}

\begin{abstract}
Materials exhibit various wave-like excitations, among which phonons (lattice vibrations) and magnons (oscillations in ferromagnetic ordering) hold significant promise for future nanoscale technologies.
Exploring the interaction between these excitations may pave the way for innovative devices that leverage their complementary strengths.
This article presents a set-up designed to amplify an incoming phononic current, potentially enhancing the phonon lifetime.
The set-up consists of a nonmagnetic and ferromagnetic insulator.
The ferromagnet is polarized opposite to the external magnetic field with spin-orbit torque, which allows for negative-energy magnons.
Phonons that are incoming from the nonmagnetic side will interact with the negative-energy magnons via magnetoelastic coupling.
The reflected phonon will increase in amplitude as a result of energy conservation.
This interaction between negative-energy magnons and phonons is an example of Klein scattering.
This work opens new avenues for the development of advanced devices that capitalize on the combined properties of phonons and magnons.
\end{abstract}

\maketitle

\section{\label{sec:Introduction}Introduction}

\begin{figure}
    \centering

\tikzset{every picture/.style={line width=0.75pt}} 

\begin{tikzpicture}[x=0.46pt,y=0.46pt,yscale=-1,xscale=1]

\draw  [draw opacity=0][fill={rgb, 255:red, 245; green, 166; blue, 35 }  ,fill opacity=0.5 ] (499.06,175.14) -- (467.42,161.1) -- (497.12,75.76) -- (538.56,61.62) -- cycle ;
\draw  [draw opacity=0][fill={rgb, 255:red, 245; green, 166; blue, 35 }  ,fill opacity=0.5 ] (467.68,161.07) -- (467.51,195.66) -- (498.72,209.54) -- (498.89,174.95) -- cycle ;
\draw  [draw opacity=0][fill={rgb, 255:red, 245; green, 166; blue, 35 }  ,fill opacity=0.5 ] (331.64,76.1) -- (496.95,76.1) -- (466.87,161.32) -- (301.56,161.32) -- cycle ;
\draw  [draw opacity=0][fill={rgb, 255:red, 80; green, 227; blue, 194 }  ,fill opacity=0.5 ] (165.74,76.1) -- (331.64,76.1) -- (301.46,161.85) -- (135.55,161.85) -- cycle ;
\draw [color={rgb, 255:red, 126; green, 211; blue, 33 }  ,draw opacity=1 ]   (292.1,146.18) .. controls (300.06,132.45) and (300.98,153.66) .. (307.16,143.84) .. controls (313.33,134.01) and (313.77,153.91) .. (321.29,142.56) ;
\draw [color={rgb, 255:red, 0; green, 0; blue, 0 }  ,draw opacity=1 ]   (204.87,144.71) .. controls (213.3,131.22) and (213.49,152.44) .. (220,142.8) .. controls (226.52,133.15) and (226.26,153.05) .. (234.18,141.93) ;
\draw [color={rgb, 255:red, 0; green, 0; blue, 0 }  ,draw opacity=1 ]   (263.53,144.93) .. controls (255.34,131.31) and (254.77,152.53) .. (248.42,142.79) .. controls (242.07,133.05) and (241.96,152.94) .. (234.24,141.7) ;
\draw [color={rgb, 255:red, 0; green, 0; blue, 0 }  ,draw opacity=1 ]   (263.86,144.33) .. controls (271.82,130.6) and (272.74,151.81) .. (278.92,141.98) .. controls (285.09,132.16) and (285.53,152.05) .. (293.05,140.71) ;
\draw  [draw opacity=0][fill={rgb, 255:red, 80; green, 227; blue, 194 }  ,fill opacity=0.5 ] (135.21,62.28) -- (165.55,75.63) -- (135.63,161.61) -- (95.96,175.06) -- cycle ;
\draw  [draw opacity=0][fill={rgb, 255:red, 80; green, 227; blue, 194 }  ,fill opacity=0.5 ] (136.02,161.32) -- (301.56,161.32) -- (301.56,195.74) -- (136.02,195.74) -- cycle ;
\draw  [draw opacity=0][fill={rgb, 255:red, 245; green, 166; blue, 35 }  ,fill opacity=0.5 ] (301.68,161.01) -- (467.23,161.01) -- (467.23,195.64) -- (301.68,195.64) -- cycle ;
\draw [color={rgb, 255:red, 74; green, 144; blue, 226 }  ,draw opacity=1 ]   (303.18,118.27) .. controls (296.89,135.59) and (293.76,101.4) .. (287.46,118.72) .. controls (281.16,136.04) and (278.03,101.85) .. (271.74,119.17) .. controls (268.44,124.59) and (265.33,119.68) .. (257.91,119.55) ;
\draw [shift={(256.01,119.62)}, rotate = 359.36] [color={rgb, 255:red, 74; green, 144; blue, 226 }  ,draw opacity=1 ][line width=0.75]    (9.84,-2.96) .. controls (6.25,-1.25) and (2.97,-0.27) .. (0,0) .. controls (2.97,0.27) and (6.25,1.26) .. (9.84,2.96)   ;
\draw [color={rgb, 255:red, 208; green, 2; blue, 27 }  ,draw opacity=1 ]   (332.43,97.34) .. controls (338.87,80.27) and (341.7,114.57) .. (348.15,97.5) .. controls (354.6,80.43) and (357.43,114.73) .. (363.88,97.66) .. controls (367.22,92.37) and (370.28,97.4) .. (377.71,97.81) ;
\draw [shift={(379.6,97.81)}, rotate = 181.56] [color={rgb, 255:red, 208; green, 2; blue, 27 }  ,draw opacity=1 ][line width=0.75]    (9.84,-2.96) .. controls (6.25,-1.25) and (2.97,-0.27) .. (0,0) .. controls (2.97,0.27) and (6.25,1.26) .. (9.84,2.96)   ;
\draw    (143.91,155.61) -- (174.8,155.61) ;
\draw [shift={(176.8,155.61)}, rotate = 180] [fill={rgb, 255:red, 0; green, 0; blue, 0 }  ][line width=0.08]  [draw opacity=0] (8.4,-2.1) -- (0,0) -- (8.4,2.1) -- cycle    ;
\draw    (143.91,155.61) -- (153.46,129.18) ;
\draw [shift={(154.13,127.29)}, rotate = 109.85] [fill={rgb, 255:red, 0; green, 0; blue, 0 }  ][line width=0.08]  [draw opacity=0] (7.2,-1.8) -- (0,0) -- (7.2,1.8) -- cycle    ;
\draw  [draw opacity=0][fill={rgb, 255:red, 80; green, 227; blue, 194 }  ,fill opacity=0.4 ] (83.69,175) -- (88.95,175) -- (88.95,209.58) -- (83.69,209.58) -- cycle ;
\draw  [draw opacity=0][fill={rgb, 255:red, 80; green, 227; blue, 194 }  ,fill opacity=0.3 ] (71.64,175.5) -- (76.9,175.5) -- (76.9,209.47) -- (71.64,209.47) -- cycle ;
\draw  [draw opacity=0][fill={rgb, 255:red, 80; green, 227; blue, 194 }  ,fill opacity=0.2 ] (58.93,175) -- (64.2,175) -- (64.2,209.58) -- (58.93,209.58) -- cycle ;
\draw  [draw opacity=0][fill={rgb, 255:red, 80; green, 227; blue, 194 }  ,fill opacity=0.1 ] (46.73,175) -- (52.11,175) -- (52.11,209.58) -- (46.73,209.58) -- cycle ;
\draw  [draw opacity=0][fill={rgb, 255:red, 80; green, 227; blue, 194 }  ,fill opacity=0.4 ] (123.12,62.25) -- (128.37,62.25) -- (88.95,175) -- (83.7,175) -- cycle ;
\draw  [draw opacity=0][fill={rgb, 255:red, 80; green, 227; blue, 194 }  ,fill opacity=0.3 ] (111.23,62.25) -- (116.45,62.25) -- (76.86,175.5) -- (71.64,175.5) -- cycle ;
\draw  [draw opacity=0][fill={rgb, 255:red, 80; green, 227; blue, 194 }  ,fill opacity=0.2 ] (98.37,62.25) -- (103.62,62.25) -- (64.2,175) -- (58.95,175) -- cycle ;
\draw  [draw opacity=0][fill={rgb, 255:red, 80; green, 227; blue, 194 }  ,fill opacity=0.1 ] (85.85,62.5) -- (91.44,62.5) -- (52.11,175) -- (46.51,175) -- cycle ;
\draw  [draw opacity=0][fill={rgb, 255:red, 80; green, 227; blue, 194 }  ,fill opacity=0.5 ] (136.4,161.17) -- (136.23,195.69) -- (95.92,209.32) -- (96.09,174.8) -- cycle ;
\draw    (136.02,161.32) -- (136.23,195.69) ;
\draw    (166.21,75.76) -- (136.02,161.51) ;
\draw  [draw opacity=0][fill={rgb, 255:red, 245; green, 166; blue, 35 }  ,fill opacity=0.4 ] (509.93,209.32) -- (503.53,209.32) -- (503.53,175.2) -- (509.93,175.2) -- cycle ;
\draw  [draw opacity=0][fill={rgb, 255:red, 245; green, 166; blue, 35 }  ,fill opacity=0.3 ] (522.17,209.57) -- (516.5,209.57) -- (516.5,175.45) -- (522.17,175.45) -- cycle ;
\draw  [draw opacity=0][fill={rgb, 255:red, 245; green, 166; blue, 35 }  ,fill opacity=0.2 ] (534.33,209.48) -- (529.06,209.48) -- (529.06,175.2) -- (534.33,175.2) -- cycle ;
\draw  [draw opacity=0][fill={rgb, 255:red, 245; green, 166; blue, 35 }  ,fill opacity=0.1 ] (545.55,209.57) -- (540.28,209.57) -- (540.28,175.45) -- (545.55,175.45) -- cycle ;
\draw  [draw opacity=0][fill={rgb, 255:red, 245; green, 166; blue, 35 }  ,fill opacity=0.4 ] (509.93,175.2) -- (503.65,175.2) -- (543.24,61.96) -- (549.52,61.96) -- cycle ;
\draw  [draw opacity=0][fill={rgb, 255:red, 245; green, 166; blue, 35 }  ,fill opacity=0.3 ] (522.12,175.45) -- (516.5,175.45) -- (556.18,61.95) -- (561.8,61.95) -- cycle ;
\draw  [draw opacity=0][fill={rgb, 255:red, 245; green, 166; blue, 35 }  ,fill opacity=0.2 ] (534.33,175.2) -- (528.81,175.2) -- (568.58,61.45) -- (574.09,61.45) -- cycle ;
\draw  [draw opacity=0][fill={rgb, 255:red, 245; green, 166; blue, 35 }  ,fill opacity=0.1 ] (545.51,175.45) -- (540.28,175.45) -- (580.05,61.7) -- (585.28,61.7) -- cycle ;
\draw    (467.42,161.06) -- (467.51,195.66) ;
\draw    (497.12,75.76) -- (467.42,161.1) ;
\draw  [color={rgb, 255:red, 0; green, 0; blue, 0 }  ,draw opacity=1 ][fill={rgb, 255:red, 74; green, 144; blue, 226 }  ,fill opacity=1 ] (286.65,142.72) .. controls (286.65,139.69) and (289.35,137.24) .. (292.68,137.24) .. controls (296,137.24) and (298.7,139.69) .. (298.7,142.72) .. controls (298.7,145.75) and (296,148.2) .. (292.68,148.2) .. controls (289.35,148.2) and (286.65,145.75) .. (286.65,142.72) -- cycle ;
\draw  [color={rgb, 255:red, 0; green, 0; blue, 0 }  ,draw opacity=1 ][fill={rgb, 255:red, 74; green, 144; blue, 226 }  ,fill opacity=1 ] (257.64,142.72) .. controls (257.64,139.69) and (260.33,137.24) .. (263.66,137.24) .. controls (266.99,137.24) and (269.68,139.69) .. (269.68,142.72) .. controls (269.68,145.75) and (266.99,148.2) .. (263.66,148.2) .. controls (260.33,148.2) and (257.64,145.75) .. (257.64,142.72) -- cycle ;
\draw  [color={rgb, 255:red, 0; green, 0; blue, 0 }  ,draw opacity=1 ][fill={rgb, 255:red, 74; green, 144; blue, 226 }  ,fill opacity=1 ] (228.18,143.12) .. controls (228.18,140.09) and (230.87,137.64) .. (234.2,137.64) .. controls (237.53,137.64) and (240.22,140.09) .. (240.22,143.12) .. controls (240.22,146.15) and (237.53,148.6) .. (234.2,148.6) .. controls (230.87,148.6) and (228.18,146.15) .. (228.18,143.12) -- cycle ;
\draw  [color={rgb, 255:red, 0; green, 0; blue, 0 }  ,draw opacity=1 ][fill={rgb, 255:red, 74; green, 144; blue, 226 }  ,fill opacity=1 ] (198.72,143.52) .. controls (198.72,140.49) and (201.42,138.04) .. (204.74,138.04) .. controls (208.07,138.04) and (210.77,140.49) .. (210.77,143.52) .. controls (210.77,146.55) and (208.07,149) .. (204.74,149) .. controls (201.42,149) and (198.72,146.55) .. (198.72,143.52) -- cycle ;
\draw [color={rgb, 255:red, 0; green, 0; blue, 0 }  ,draw opacity=1 ]   (409.38,140.56) .. controls (400.92,154.14) and (400.73,132.78) .. (394.2,142.48) .. controls (387.66,152.19) and (387.92,132.17) .. (379.98,143.36) ;
\draw [color={rgb, 255:red, 0; green, 0; blue, 0 }  ,draw opacity=1 ]   (350.53,140.33) .. controls (358.75,154.04) and (359.32,132.69) .. (365.69,142.49) .. controls (372.06,152.3) and (372.17,132.27) .. (379.91,143.59) ;
\draw [color={rgb, 255:red, 0; green, 0; blue, 0 }  ,draw opacity=1 ]   (350.2,140.94) .. controls (342.21,154.77) and (341.29,133.41) .. (335.09,143.31) .. controls (328.9,153.2) and (328.46,133.16) .. (320.91,144.59) ;
\draw [color={rgb, 255:red, 74; green, 144; blue, 226 }  ,draw opacity=1 ]   (327.88,119.53) .. controls (334.34,102.59) and (337.55,136.04) .. (344.02,119.09) .. controls (350.48,102.15) and (353.69,135.6) .. (360.16,118.65) .. controls (363.54,113.36) and (366.73,118.16) .. (374.36,118.28) ;
\draw [shift={(376.3,118.21)}, rotate = 179.39] [color={rgb, 255:red, 74; green, 144; blue, 226 }  ,draw opacity=1 ][line width=0.75]    (9.84,-2.96) .. controls (6.25,-1.25) and (2.97,-0.27) .. (0,0) .. controls (2.97,0.27) and (6.25,1.26) .. (9.84,2.96)   ;
\draw  [color={rgb, 255:red, 0; green, 0; blue, 0 }  ,draw opacity=1 ][fill={rgb, 255:red, 74; green, 144; blue, 226 }  ,fill opacity=1 ] (327.34,142.56) .. controls (327.34,145.61) and (324.63,148.08) .. (321.29,148.08) .. controls (317.96,148.08) and (315.25,145.61) .. (315.25,142.56) .. controls (315.25,139.52) and (317.96,137.05) .. (321.29,137.05) .. controls (324.63,137.05) and (327.34,139.52) .. (327.34,142.56) -- cycle ;
\draw  [color={rgb, 255:red, 0; green, 0; blue, 0 }  ,draw opacity=1 ][fill={rgb, 255:red, 74; green, 144; blue, 226 }  ,fill opacity=1 ] (356.45,142.56) .. controls (356.45,145.61) and (353.74,148.08) .. (350.4,148.08) .. controls (347.07,148.08) and (344.36,145.61) .. (344.36,142.56) .. controls (344.36,139.52) and (347.07,137.05) .. (350.4,137.05) .. controls (353.74,137.05) and (356.45,139.52) .. (356.45,142.56) -- cycle ;
\draw  [color={rgb, 255:red, 0; green, 0; blue, 0 }  ,draw opacity=1 ][fill={rgb, 255:red, 74; green, 144; blue, 226 }  ,fill opacity=1 ] (386,142.16) .. controls (386,145.21) and (383.29,147.68) .. (379.96,147.68) .. controls (376.62,147.68) and (373.91,145.21) .. (373.91,142.16) .. controls (373.91,139.11) and (376.62,136.64) .. (379.96,136.64) .. controls (383.29,136.64) and (386,139.11) .. (386,142.16) -- cycle ;
\draw  [color={rgb, 255:red, 0; green, 0; blue, 0 }  ,draw opacity=1 ][fill={rgb, 255:red, 74; green, 144; blue, 226 }  ,fill opacity=1 ] (415.55,141.76) .. controls (415.55,144.8) and (412.84,147.27) .. (409.51,147.27) .. controls (406.17,147.27) and (403.46,144.8) .. (403.46,141.76) .. controls (403.46,138.71) and (406.17,136.24) .. (409.51,136.24) .. controls (412.84,136.24) and (415.55,138.71) .. (415.55,141.76) -- cycle ;
\draw    (415.3,125.2) -- (460.79,124.92) ;
\draw [shift={(462.79,124.91)}, rotate = 179.65] [color={rgb, 255:red, 0; green, 0; blue, 0 }  ][line width=0.75]    (10.93,-3.29) .. controls (6.95,-1.4) and (3.31,-0.3) .. (0,0) .. controls (3.31,0.3) and (6.95,1.4) .. (10.93,3.29)   ;
\draw [color={rgb, 255:red, 208; green, 2; blue, 27 }  ,draw opacity=1 ]   (470.72,106.5) -- (424.8,106.69) ;
\draw [shift={(422.8,106.7)}, rotate = 359.76] [color={rgb, 255:red, 208; green, 2; blue, 27 }  ,draw opacity=1 ][line width=0.75]    (10.93,-3.29) .. controls (6.95,-1.4) and (3.31,-0.3) .. (0,0) .. controls (3.31,0.3) and (6.95,1.4) .. (10.93,3.29)   ;
\draw [color={rgb, 255:red, 74; green, 144; blue, 226 }  ,draw opacity=1 ]   (264.68,97.13) .. controls (271.14,80.19) and (274.35,113.64) .. (280.82,96.69) .. controls (287.28,79.75) and (290.49,113.2) .. (296.96,96.25) .. controls (300.34,90.96) and (303.53,95.76) .. (311.16,95.88) ;
\draw [shift={(313.1,95.81)}, rotate = 179.39] [color={rgb, 255:red, 74; green, 144; blue, 226 }  ,draw opacity=1 ][line width=0.75]    (9.84,-2.96) .. controls (6.25,-1.25) and (2.97,-0.27) .. (0,0) .. controls (2.97,0.27) and (6.25,1.26) .. (9.84,2.96)   ;

\draw (142.45,171.06) node [anchor=north west][inner sep=0.75pt]  [font=\small] [align=left] {NM};
\draw (435.08,170) node [anchor=north west][inner sep=0.75pt]  [font=\small] [align=left] {FM};
\draw (154.27,115.43) node [anchor=north west][inner sep=0.75pt]  [font=\footnotesize]  {$z$};
\draw (166.52,139.08) node [anchor=north west][inner sep=0.75pt]  [font=\footnotesize]  {$x$};
\draw (433.58,129.54) node [anchor=north west][inner sep=0.75pt]    {$B$};
\draw (440.88,79.88) node [anchor=north west][inner sep=0.75pt]  [color={rgb, 255:red, 208; green, 2; blue, 27 }  ,opacity=1 ]  {$\vec{m}_{R}$};

\end{tikzpicture}
    \caption{A schematic of a nonmagnetic (NM) material coupled to a ferromagnet (FM) that is oppositely aligned with its external magnetic field. There are two wavelike excitations, lattice vibrations (phonons) and oscillations in the ferromagnetic ordering (magnons). Phonons (magnons) are represented by the blue (red) arrow. There is one incoming phonon, which scatters into a reflected phonon and a transmitted magnon and phonon.}
    \label{fig:set-up1}
\end{figure}
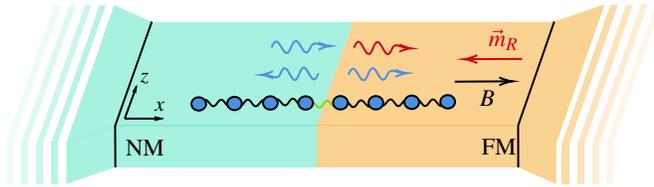

Magnetic insulators exhibit two types of wavelike excitations: lattice vibrations or phonons and oscillations of ferromagnetic ordering or magnons. 
Both have been the focus of research for nanoscale devices aimed at efficiently storing and manipulating information\cite{flebus_2024_2024,xie_brief_2023}.
Integrating both phonons and magnons offers complementary advantages that can enhance device performance. 
Phonons, with their long coherence times, are ideal for stable, low-energy information processing. 
Magnons, on the other hand, enable high-speed operations due to their wide frequency range and can facilitate rapid information transfer. 
By leveraging the strengths of both excitations, it is possible to develop hybrid systems that capitalize on the stability of phonons and the speed of magnons, leading to more efficient and versatile nanoscale devices.

The interaction between lattice vibrations and magnetic dynamics was first investigated over fifty years ago by Kittel\cite{kittel_relaxation_1953,kittel_interaction_1958}, Abraham \cite{kittel_relaxation_1953}, Kaganov, and Tsukernik \cite{kaganov1959phenomenological}.
These interactions are crucial in several relaxation processes\cite{kittel_relaxation_1953,ruckriegel_magnetoelastic_2014} and in the dissipation of heat, such as in the spin Seebeck effect\cite{uchida_spin_2010,uchida_long-range_2011,kikkawa_magnon_2016,flebus_magnon-polaron_2017}. 
Additionally, spin pumping has been achieved by injecting coherent elastic waves\cite{kamra_coherent_2015,uchida_long-range_2011,weiler_spin_2012}, demonstrating the interplay between magnetic and elastic waves.
More recent advances include phonon-magnon coupling in a layered antiferromagnet\cite{lyons_acoustically_2023}.
Reviews\cite{bozhko_magnon-phonon_2020,li_advances_2021} on the subject, together with this special issue on ``Phonon-Magnon Interactions'' are a testament to the growing significance of this research area even after fifty years.

To reliably use phonons and magnons in devices, we need reliable transfer of information by them.
Both types of excitation are compromised by scattering processes, which causes their amplitude to diminish over time.
The necessity to amplify phononic currents to effectively compound nanoscale devices has been apparent, and its remedies have been studied in several articles.
In particular, phonon amplifiers or lasers have been proposed that drive the phononic field using photons\cite{jing_mathcalpt-symmetric_2014,poshakinskiy_multiple_2016}, electrons\cite{christensen_parity-time_2016,hackett_s-band_2024}, or temperature gradients\cite{aspelmeyer_cavity_2014}.

In this article, we investigate a hybrid structure of a nonmagnetic material and a ferromagnet.
This set-up has been considered in several references\cite{kamra_coherent_2015,kamra_actuation_2014,weiler_spin_2012,uchida_long-range_2011}.
These articles investigate the magnetization dynamics after the injection of phonons.
Our set-up differs since the ferromagnet is stabilized opposite to the external magnetic field due to driving, see Fig. \ref{fig:set-up1}.
This driving can be achieved through spin-orbit torque, which causes the configuration to be kept in this non-equilibrium state.
We are interested in the phononic response to the steady-state non-equilibrium magnet.
We compute the energy flux using scattering theory.
We show that for an incoming phonon current from the nonmagnetic insulator, the reflected phonon current is amplified for frequencies close to the ferromagnetic resonance.
Microscopically, the excitation of magnons in the non-equilibrium state lowers the total energy of the magnet; therefore, the hybridized phonons and magnons due to magnetoelastic coupling also lower the energy of the magnet.
Thus, the incoming phonon then scatters into lower energy states which in turn gives a greater reflected elastic energy flux.
A similar effect for magnons has been shown in Ref. [\onlinecite{Joren}].
The spin-orbit torque acts as a reservoir that provides the ferromagnet with sufficient energy for the amplification process.
Our set-up differs from the other phononic amplifiers in its simplicity and in the tunability of the amplified phonon frequency by changing the external magnetic field.
Our proposed set-up may pave the way for phonon amplification in devices that utilize both magnons and phonons.

The article is constructed as follows.
We introduce the set-up and the equations with which we model the system in Section \ref{sec:set-up and Theory}.
In Section \ref{sec:Complex Field equations}, we rewrite the equations of motion in terms of complex fields and solve for a plane wave ansatz.
Then in Section \ref{sec:Energy Flux}, we compute the energy flux and use scattering theory techniques to find the phonon current between the nonmagnetic insulator and the magnetic insulator.
This will lead to reflection and transmission for an initial phonon current.
We summarize and give an overview of possible future avenues in Section \ref{sec:Summary and Outlook}.

\section{\label{sec:set-up and Theory}set-up and Theory}

The set-up consists of two layers; one of the layers is a nonmagnetic (NM) material and the other layer is a ferromagnetic (FM) material, as in Fig. \ref{fig:set-up1}.
There is also an external magnetic field given by $\vec B=\mu_0 H \hat x$ with $\mu_0$ the vacuum permeability and $H$ the magnetic field strength.
The ferromagnetic material is stabilized to point against the effective field by injecting spin-angular momentum.
This is achieved by adding a heavy metal layer to the ferromagnet and using the inverse spin hall effect to exert spin-orbit torque on the spins in the ferromagnet\cite{miron_perpendicular_2011,liu_current-induced_2012,kurebayashi_antidamping_2014, manchon_new_2015}.
In this section, we give the Hamiltonian density for the magnetization ($\vec M$) and the displacement field ($\vec R$) of this set-up.
The ferromagnet is polarized oppositely to the external field, thus its magnetization is simplified to $M_z,M_y\ll M_s$ and $M_x\sim -M_s$ after linearization with $M_s$ the saturation magnetization.
The Hamiltonian density is given by a magnetic, phononic and magnetoelastic coupling contribution.
These contributions are given in Kittel's paper\cite{kittel_interaction_1958}.
The magnetic contribution consists of an exchange interaction and a Zeeman interaction

\begin{equation}
    \Ha_M=\frac{A}{M_s^2}\left[(\nabla M_y)^2+(\nabla M_z)^2\right]-\frac{\omega_0}{2\gamma M_s} \left(M_z^2+M_y^2\right).
\end{equation}

\noindent Here we have defined $\omega_0=\gamma \mu_0 H$ as the ferromagnetic resonance frequency with $\gamma$ the gyromagnetic ratio and $A$ is the spin stiffness.
The phononic contribution is given by

\begin{equation}
    \Ha_{E}=\half \rho (\partial_t \vec{R})^2+ \lambda\left(\sum_i S_{ii}\right)^2+\mu_F\sum_{ij} S_{ij}^2.
\end{equation}

\noindent The components of the strain tensor are given by $S_{ij}=\half(\partial_{j}R_i+\partial_{i}R_j)$ in terms of the displacement field $R_i$ with $i,j\in\{x,y,z\}$. 
We define $\rho$ as the mass density and Lam\'e's constants $\lambda$ and $\mu_F$.
For cubic symmetry, the magnetoelastic coupling is modelled by

\begin{equation}
    \Ha_{MEC}=\frac{b_1}{M_s^2}\sum_iM_i^2S_{ii}^2+\frac{b_2}{M_s^2}\sum_{i\neq j} M_iM_jS_{ij}
\end{equation}

\noindent with $b_1$ and $b_2$ phenomenological coupling constants.
The sum of the contributions is the complete Hamiltonian density $\Ha=\Ha_M+\Ha_E+\Ha_{MEC}$ from which the the equations of motion are derived,

\begin{subequations}
\begin{eqnarray}\label{eq:config2real1}
    \dot M_z&&=-\omega_0 M_y-J\partial_x^2 M_y-b_2\gamma\partial_x R_y, \\
    \dot M_y&&=\omega_0 M_z+J\partial_x^2 M_z+b_2\gamma\partial_x R_z, \\
    \rho \ddot R_z&&=\mu_F \partial_x^2 R_z-\frac{b_2}{M_s}\partial_x M_z, \\
    \rho \ddot R_y&&=\mu_F \partial_x^2 R_y-\frac{b_2}{M_s}\partial_x M_y, \label{eq:config2real2}
\end{eqnarray}    
\end{subequations}

\noindent where $J=2\gamma A/M_s$. 
Only the $x$ direction is relevant for the following analysis, thus only derivatives in $x$ are taken into account.
In the following section, we express the equations of motion in a convenient complex basis.
We are ultimately interested in the energy flux, which satisfies $\partial_t\Ha=-\partial_xF$.
Using the equations of motion, we find that

\begin{eqnarray}
    F&&=-\frac{2A}{M_s^2}\left(\partial_t M_z\partial_x M_z+\partial_tM_y\partial_xM_y\right) \nonumber\\
    &&-\mu_F\left(\partial_tR_z\partial_x R_z+\partial_tR_y\partial_x R_y\right) \nonumber\\
    &&+\frac{b_2}{M_s}\left(M_z\partial_tR_z+M_y\partial_tR_y\right).
\end{eqnarray}

\noindent Note that we have explicitly written down the equation for the ferromagnetic material. The equations of motion for the nonmagnetic material are similar but without the contribution due to the magnetization and magnetoelastic coupling. Throughout this work, we take the mass density of ferromagnetic and nonmagnetic insulator to be the same, for convenience. Generalization to the more general case of unequal mass densities is straightforward. 


\section{\label{sec:Complex Field equations}Phonons and magnetoelastic waves}

\begin{figure*}
    \centering
    \includegraphics[width=510px]{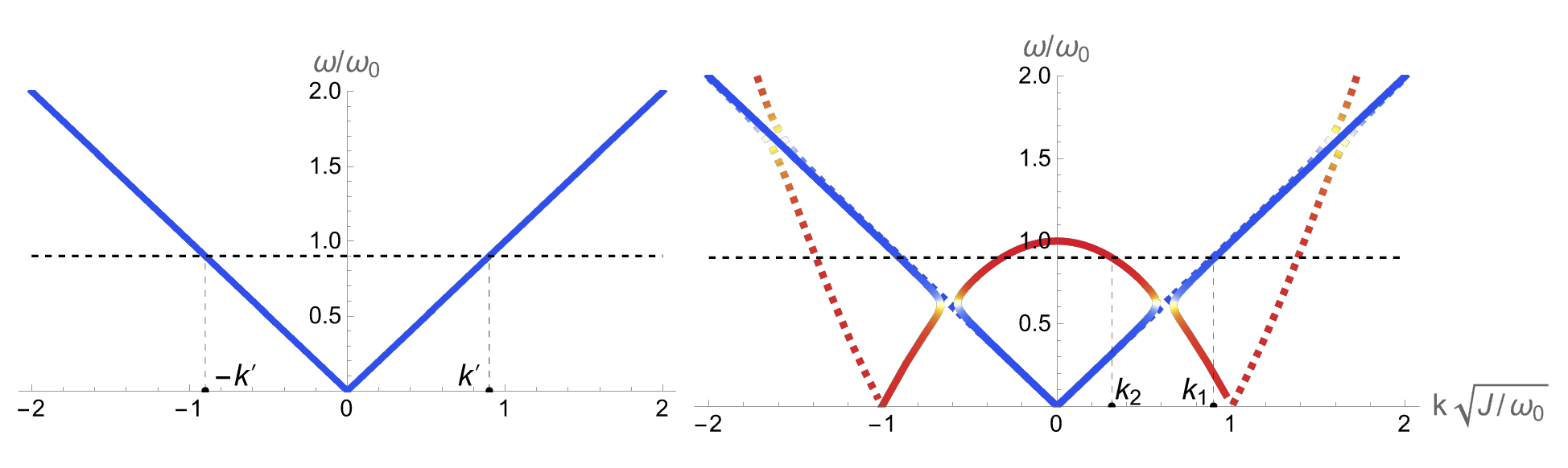}
    \caption{The dispersion relations in the nonmagnetic material (left) and the ferromagnetic (right) with the normalised quantities $\omega/\omega_0$ and $k\sqrt{J/\omega_0}$. The colours indicate the degree of magnonic (red) or phononic (blue) signature. The coupling $b_2$ is taken to be $0.1M_s\omega_0/\gamma$. The horizontal dashed line is a frequency of interest because at this frequency the negative-energy magnons in the ferromagnet couple to the phonons in the nonmagnetic insulator.  At this frequency, there are two wave vector solutions, $k_1$ and $k_2$, in the right figure.}
    \label{fig:DispersionConfig2}
\end{figure*}

In this section, we bring the equations of motion Eq. (\ref{eq:config2real1}-\ref{eq:config2real2}) into a complex basis.
This will turn out to be a convenient choice for our purposes.
The dimensionless complex fields for the magnetic degrees of freedom and the phononic degrees of freedom are given by

\begin{eqnarray}
    \phi_m=\frac{M_x+iM_y}{M_s}, && \phi_p=\frac{R_z+i R_y}{a}.
\end{eqnarray}

\noindent Here $a$ is the lattice constant.
We substitute the complex fields in Eq. (\ref{eq:config2real1}-\ref{eq:config2real2}), which yields the equations of motion

\begin{subequations}
\begin{eqnarray}
   i\dot \phi_m&&=-\omega_0\phi_m-J\partial_x^2\phi_m-b_2\frac{\gamma a}{M_s}\partial_x\phi_p, \\
    i\dot \phi_m^*&&=\omega_0\phi_m^*+J\partial_x^2\phi_m^*+b_2\frac{\gamma a}{M_s}\partial_x\phi_p^*, \\
    \ddot \phi_p&&=\frac{\mu_F}{\rho}\partial_x^2\phi_p-\frac{b_2}{\rho a}\partial_x\phi_m, \\
    \ddot \phi_p^*&&=\frac{\mu_F}{\rho}\partial_x^2\phi_p^*-\frac{b_2}{\rho a}\partial_x\phi_m^*.  
\end{eqnarray}
\end{subequations}

\noindent The expression for the energy flux in terms of these complex fields is

\begin{eqnarray}
        F&&=-A\left(\partial_t \phi_m\partial_x \phi_m^*+\partial_t \phi_m^*\partial_x \phi_m\right) \nonumber\\
    &&-\frac{a^2 \mu_F}{2}\left( \partial_t \phi_p\partial_x\phi_p^*+\partial_t \phi_p^*\partial_x\phi_p\right) \nonumber\\
    &&+\frac{b_2 a}{2}\left(\phi_m\partial_t \phi_p^*+\phi_m^*\partial_t \phi_p\right).
\end{eqnarray}

\noindent Consider the plane-wave ansatz

\begin{align}\label{eq:anstatz}
    \begin{pmatrix}
\phi_m \\
\phi_m^* \\
\phi_p \\
\phi_p^* 
\end{pmatrix}=
e^{ikx-i\omega t}\begin{pmatrix}
u_m \\
v_m \\
u_p \\
v_p 
\end{pmatrix}.
\end{align}

\noindent This ansatz together with the equations of motion results in a set of four equations, which is compactly written as $W (u_m,v_m,u_p,v_p)^T=0$.
The matrix $W$ has the following form

\begin{equation}
        W=\begin{pmatrix}
            \omega-\omega_m(k) & 0 & i\frac{b_2 \gamma a}{M_s}k & 0 \\
            0 & \omega+\omega_m(k) & 0 & -i\frac{b_2 \gamma a}{M_s}k \\
            i\frac{b_2 \rho}{a}k & 0 & \omega_p^2-\omega^2 & 0 \\
            0 & i\frac{b_2 \rho}{a}k & 0 & \omega_p^2-\omega^2 \\
        \end{pmatrix}.
\end{equation}

\noindent The magnon and phonon dispersions in the absence of magnetoelastic coupling are $\omega_m(k)=J k^2-\omega_0$ and $\omega_p(k)=k\sqrt{\mu_F/\rho}$.
Note that for $k=0$, we have that $\omega_m =-\omega_0$ due to the fact that we consider the out-of-equilibrium situation of the magnet that is driven opposite to the field.
The matrix above describes two sets of coupled equations that physically correspond to coupled magnon-phonon excitation, dubbed magnetoelastic waves.
The two solutions correspond to two different polarizations. 
For the equation $W (u_m,v_m,u_p,v_p)^T=0$ to have non-trivial solutions, we require $det[W]=0$.
This gives us 6 unwieldy expressions for $\omega$, two of which are imaginary and unphysical, and the other four solutions are plotted in the right panel of Fig. \ref{fig:DispersionConfig2}.
Here, we see four dispersion relations that in the absence of magnetoelastic coupling correspond to $\pm\omega_m(k)$ and $\pm\omega_p(k)$.
Because of magnetoelastic coupling, the dispersion relations hybridize.
The hybridization causes a smooth transition between phonon and magnon dominated regimes.
This hybridization is indicated by color in the figure: red regions represent predominantly magnonic character, while blue regions correspond to predominantly phononic character.
Additionally, the dispersion relations are distinguished by solid and dashed lines, which correspond to different polarization of the magnetoelastic waves.
The dashed red line is associated with positive energy magnons.
We assume a positive gyromagnetic ratio such that these magnons are clockwise circularly polarized.
The clockwise circularly polarized magnons can only hybridize with clockwise circularly polarized phonons, given by the dashed blue line.
The solid red line is associated with counterclockwise circularly polarized negative energy magnons, or antimagnons.
These magnons, typically absent in conventional systems, become physical due to the oppositely aligned effective field.
Antimagnons carry negative energy\cite{harms_antimagnonics_2024}, and hybridize with counterclockwise circularly polarized phonons (solid blue line).
This hybridization produces counterclockwise polarized magnetoelastic waves, which have lower energy than uncoupled phonons due to the contribution of the antimagnon’s negative energy.

The left part of Fig. \ref{fig:DispersionConfig2} represents the phonon dispersion in the nonmagnetic layer.
Here, too, we find two differently circularly polarized phonons.
We consider the set-up as in Fig. \ref{fig:set-up1} with one incoming phonon from the nonmagnetic layer.
First, consider a clockwise polarized phonon incoming from the left.
We consider frequencies between $0$ and $\omega_0$ because for these frequencies the negative-energy magnons in the ferromagnetic insulator couple to the phonons in the non-magnetic insulator. 
An example of such a frequency is the horizontal dashed line in Fig. \ref{fig:DispersionConfig2}.
This clockwise polarized phonon scatters into a reflected phonon and two clockwise polarized magnetoelastic waves.
This scattering process is known in the literature\cite{kamra_actuation_2014}.
Now, consider an incoming counterclockwise polarized phonon with a frequency given by the horizontal line in Fig. \ref{fig:DispersionConfig2}.
The scattering process results in a reflected phonon ($-k'$) and two transmitted counterclockwise polarized magnetoelstic waves with wavevectors $k_1$ and $k_2$.
As discussed before, these magnetoelastic waves are low in energy due to the antimagnon contribution.
For the scattering process to conserve energy, it has to increase the amplitude of the reflected phonon.
Therefore this scattering process enhances phonon reflection, which is the main result of this article.
The counterclockwise polarized magnetoelastic waves are described by the components ($v_m$, $v_p$).
The coefficients $v_m$ and $v_p$ are the magnon and phonon components of the plane-wave ansatz, respectively, defined in Eq. (\ref{eq:anstatz}).
They have the following form

\begin{subequations}
\begin{eqnarray}
    v_m&&=-N\frac{i a \gamma  k b_2 }{M (\omega +\omega_m(k))}, \\
    v_p&&=N.
\end{eqnarray}
\end{subequations}

\noindent The coefficients have an overall normalisation factor $N$.
The energy flux is given in terms of the components ($v_m$, $v_p$) as

\begin{eqnarray}
\bar F&&=2A \omega k |v_{m}|^2 +\frac{a^2 \mu_F  \omega k}{2} |v_p|^2 \nonumber\\
    &&+a b_2  \omega  \Im[v_{p}  v_{m}^*].
\end{eqnarray}

\noindent Here, the energy flux has been averaged over time.
The scattering process described above together with boundary conditions that are given in the following section allows us to compute the energy fluxes after the scattering process.

\section{\label{sec:Energy Flux}Energy Flux}

In this section, we compute the scattering of an incoming counterclockwise polarized phonon in the nonmagnetic layer as described above.
The composition of the excitations relevant in the scattering process is given by

\begin{subequations}
\begin{eqnarray}\label{eq:NMwaves}
    \Psi_{x< 0}&&=e^{-i\omega t}\left(e^{ik'x}+re^{-ik'x}\right)\begin{pmatrix}
0 \\
v_p'
\end{pmatrix}, \\ \label{eq:FMwaves}
\Psi_{x> 0}&&=t_1e^{ik_1x-i\omega t}\begin{pmatrix}
v_{m1} \\
v_{p1} 
\end{pmatrix}+t_2e^{ik_2x-i\omega t}\begin{pmatrix}
v_{m2} \\
v_{p2}
\end{pmatrix} .
\end{eqnarray}
\end{subequations}

\noindent To impose energy conservation onto the scattering process, we use boundary conditions (BC).\
Define the interface between the nonmagnetic and ferromagnetic layer to be at $x=0$.
First, the phononic field should be continuous between the layers.

\begin{equation}
    \vec R|_{x\uparrow 0}=\vec R|_{x\downarrow 0}.
\end{equation}

\noindent Secondly, we impose that no magnetic angular momentum should pass into the nonmagnetic insulator.

\begin{equation}
    \partial_x \vec M|_{x\downarrow 0}=0.
\end{equation}

\noindent This condition is derived by integrating Eq. (\ref{eq:config2real1}) over the interface.
For the last BC, we impose that the energy flux is continuous which in turn enforces energy conservation:

\begin{equation}
    F|_{x\uparrow 0}=F|_{x\downarrow 0}.
\end{equation}

\noindent The momenta are found by inverting the dispersion relation.
The momentum for phonon is $k'(\omega)=\omega\sqrt{\rho/\mu_F}$.
The expressions for the momenta of the magnetoelastic waves $k_1$ and $k_2$ are complicated and not presented here.
Imposing the BC onto the scattering state gives us the following equations to solve for 

\begin{subequations}
\begin{eqnarray}
    \sum_{i=1,2}k_i t_i v_{mi} &&=0, \\
    1+r&&= \sum_{i=1,2} t_iv_{pi}, \\
    ik'(1-r)&&=\sum_{i=1,2}ik_it_iv_{pi}-\frac{b_2}{a\mu_F}\left(\sum_{i=1,2}t_iv_{mi}\right).
\end{eqnarray}
\end{subequations}

\begin{figure}
    \centering
    \includegraphics[width=245px]{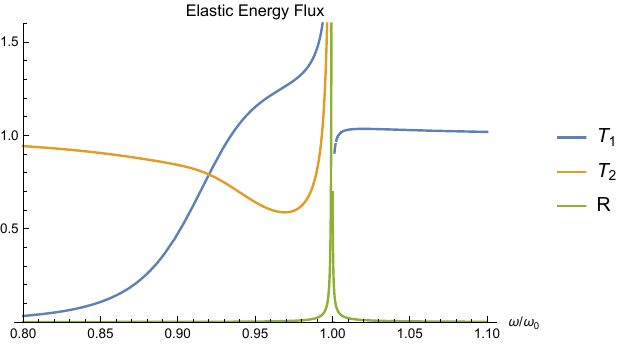}
    \caption{The elastic energy flux for the transmissions and the reflected plane wave in the second configuration. Here we have taken $J=0.1a^2\omega_0$ and $b_2=0.1M_s\omega_0/\gamma$.}
    \label{fig:ElasticEnergyFluxconfig2}
\end{figure}

\noindent Unfortunately the solutions to the coefficients $r$, $t_1$ and $t_2$ are unwieldy and not presented here.
The solutions to the coefficients allow us to compute the energy fluxes.
We define the reflected energy flux as $R=|\Bar F_r/\Bar F_i|=|r|^2$, where $\Bar F_r$ is computed using the reflected plane wave and $\Bar F_i$ is the flux computed with the incident plane wave in Eq. (\ref{eq:NMwaves}). 
The total transmitted energy flux is defined as $T_{tot}=|\Bar F_T/\Bar F_i|$, where $\Bar F_T$ is computed using the sum of the outgoing plane waves in the ferromagnet, Eq. (\ref{eq:FMwaves}).
The reflected and total transmitted energy flux conserve energy, $R+T_{tot}=1$.
Separate channels have individual elastic energy fluxes $T_1=|\Bar F_{t_1}/\Bar F_i|$ and $T_2=|\Bar F_{t_2}/\Bar F_i|$ which are computed with the corresponding plane waves in Eq. (\ref{eq:FMwaves}).
Individual fluxes as a function of frequency are presented in Fig. \ref{fig:ElasticEnergyFluxconfig2}.
There are two defining features: the switching between channels in which $T_1-T_2$ changes sign.
This occurs at approximately the same frequency as the avoided crossing in the dispersion. 
The fluxes for frequencies close to the ferromagnetic resonance become greater than one. 
This corresponds to the enhanced reflection that is the focus of this article. 
In Fig. \ref{fig:ElasticEnergyFluxconfig2}, the fluxes not only exceed one but also diverge for frequencies close to the ferromagnetic resonance.
We expect that a non-linear analysis would lead to a phononic response that has a well-defined maximum. 
This is, however, beyond the scope of this paper.

The phonon amplification is a qualitative effect and therefore does not dependent on material parameters in the sense that it always exists at some frequency in our set-up, provided the magnetic is driven to point opposite to the effective field.
However, the effect can vary greatly in strength for different materials.
The most suitable parameters should be large magnetoelastic coupling ($b_2$) and low Gilbert damping to efficiently polarize the ferromagnetic oppositely to the external magnetic field.
Yttrium Iron Garnet (YIG) has an extremely low damping\cite{gurevich_magnetization_2020}, however, it also has a relatively low magnetoelastic coupling\cite{gurevich_magnetization_2020}.
This is partially remedied by the fact that YIG also has a low spin stiffness, resulting in a large magnon density of states close to the frequency range of hybridization between magnons and phonons.
The other layer has to be nonmagnetic and it has to match the ferromagnetic interface.
In the case of YIG, gadolinium gallium garnet (GGG) would be a good interfacial match and it is nonmagnetic at large temperatures.

\section{\label{sec:Summary and Outlook}Summary and Outlook}

In this article, we showed that a nonmagnetic layer coupled to a ferromagnet that is polarized against the effective field amplifies an incoming phonon current.
This is because the hybridized magnon-phonon excitation, or magnetoelastic wave, in the magnet lowers the energy of the system.
To conserve energy, the reflected phonon current must be greater than the initial phonon current.
In our simplified model, the amplification diverges for a frequency close to the ferromagnetic resonance.
If we take into account higher-order terms.
We expect the divergence to change into a well-defined maximum.

The set-up we proposed serves as an amplifier for a phononic current with frequencies close to the ferromagnetic resonance.
In addition, it would be an effective way to excite a single circularly polarized phonon current at a well-defined frequency.
The choice depends on the direction in which the ferromagnet is polarized, as long as the ferromagnet is oppositely polarized to the external magnetic field and points along the $x$ direction.
In the case that the external magnetic field points perpendicular to the $x$ direction, we find an equivalent result to Ref. [\onlinecite{kamra_coherent_2015}].
This is unsurprising considering that this configuration only allows one phonon degree of freedom to couple to the magnons.
Thus, angular momentum will not be able to transfer from the ferromagnet to the nonmagnetic layer.
Due to the conservation of angular momentum, no amplification of the phonon current is possible for this configuration.

In the current set-up, we treated the layers as if they were infinitely long.
If the ferromagnetic layer has a well-defined width, then standing waves dominate the spectrum.
If a phonon current with a particular frequency coalesces with the frequency of a standing wave, then the phonon current will be amplified as described in this article.
In this process, the standing wave itself is also amplified, which in turn amplifies the phonon current more.
This recurrence amplifies a single-frequency phonon current in a way that is similar to lasing.
To investigate this effect would require a non-linear analysis.
A similar effect with two ferromagnetic layers has been theoretically shown in Ref. [\onlinecite{Joren_Laser}].

Magnetic moments stabilized against the magnetic field experience large quantum-mechanical zero-point motion
Therefore any coupling to the magnetic moments can lead to zero-point motion in the coupled degree of freedom.
Thus, in the set-up of our article, we expect phonons to be spontaneously excited because of the coupling with magnetic moments.
A similar effect has been theoretically demonstrated in Ref. [\onlinecite{bassant_entangled_2024}] for two ferromagnets.
We also expect the phonons in the nonmagnetic insulator to be entangled with the magnetoelastic waves in the ferromagnet.

This work concerns a steady-state non-equilibrium magnet coupled to a phononic system.
To extend this article, one can investigate other wavelike excitations that can couple to magnons, like photons or plasmons\cite{costa_strongly_2023}.

\begin{acknowledgments}
This work is funded by the projects “Black holes on a chip” with project number OCENW.KLEIN.502 and “Fluid Spintronics” with project number VI.C.182.069.
Both are financed by the Dutch Research Council (NWO).
\end{acknowledgments}

\nocite{*}
\bibliography{aipsamp}

\end{document}